\documentstyle[preprint,aps]{revtex}

\def\eps{\epsilon}
\begin{document}

\title{Noise-enhanced reconstruction of attractors}

\author{Rolando Castro and Tim Sauer}
\address{Institute of Computational Sciences and Informatics \\and
Department of Mathematical Sciences \\ George Mason
University \\ Fairfax, VA 22030} 
\maketitle

\begin{abstract}
In principle, the state space of a chaotic attractor
can be partially or wholly reconstructed from interspike intervals
recorded from experiment. Under certain conditions, the quality of a
partial reconstruction, as measured by the spike train prediction error,
can be increased by adding noise to the spike creation process. This
phenomenon for chaotic systems is an analogue of stochastic resonance.
\end{abstract}

 The seminal articles \cite{PCFS} and the theorem of Takens in
\cite{takens} demonstrated the theoretical and practical possibility
of reconstructing 
the topological structure of the state space underlying an
experimental system, using the measurement of a generic
scalar or multivariate signal from the system. This possibility is
especially welcome for nonlinear systems, where the potential
exists for
extremely complicated state space attractors. A great deal
of subsequent research effort has gone into 
developing data processing techniques for the detection, analysis and
exploitation of nonlinear, and in particular chaotic, processes.

Often, a delay coordinate reconstruction of a
compact attractor from an evenly-sampled time series of measurements
can be found that 
is topologically equivalent to the attractor. The function mapping the
attractor to the reconstructed copy is called an embedding. According
to \cite{saueryc}, as long as the embedding dimension is greater than twice 
the box-counting dimension of the attractor, an embedding results
for a probability-one choice of measurement functions. Recent theoretical
work has attempted to widen the scope of dynamical data that can lead
to an embedding. It has been shown, for example, that a
similar reconstruction result holds when using spike train
data (the recorded times between firings) from a model
integrate-and-fire dynamical system with chaotic dynamics \cite{sauer}.

Attractor reconstruction can be viewed as a type of information transfer.
In the case of a topological embedding, no information is lost. The
set of states of the underlying experimental system is reproduced
exactly in the copy that is reconstructed from measured data. In other
cases, the reconstruction may be incomplete. When a chaotic signal is
fed into a threshold crossing detector, the time-delay plot of
time intervals between crossings reconstructs something akin to a
Poincar\'e section 
of the underlying chaotic attractor. The dimension is decreased by one
\cite{CS}. A  recent study
\cite{andre} of neuron models subjected to chaotic input
points out that the two-dimensional FitzHugh-Nagumo 
differential equation \cite{FHN} (hereafter referred to as FHN2) acts
as a kind of 
threshold-crossing ``filter'' for input signals, and in particular fails
to completely reconstruct the attractor which generated the input
signal. One focus of the present article is to clarify the
distinction, for attractor reconstruction purposes, between
threshold-crossing (TC) and integrate-and-fire (IF) filters. In
contrast to the fact that FHN2 acts as a TC filter, we exhibit
a variation on FHN2 which acts principally as an
IF filter instead, and does fully reconstruct the input 
attractor. This differential equation model, which we shall denote
FHN3, is a three-dimensional version of excitable FitzHugh-Nagumo-type 
dynamics.  With
appropriate parameter settings, it has the following remarkable
property: When a signal from a system attractor is added to one of
FHN3's variables, another variable undergoes a deterministic spiking
behavior whose interspike intervals, embedded as $m$-tuples, embed the
original attractor in $m$-dimensional reconstruction space.
Just as Takens' theorem guarantees that multidimensional state
space information can be condensed into a single evenly-spaced time
series, the same can be accomplished with spike timings from the FHN3 
filter.

Viewing attractor reconstruction as information transfer raises
questions about the efficiency of the transfer process, and the
possible effects of noise on this process. Surprisingly, in some
instances noise can have a beneficial effect on information
transmission, analogous to the stochastic resonance phenomenon
\cite{sr} observed for multistable potentials and excitable media.
This seemingly contradictory effect is the observed amplification 
of a filtered signal achieved when stochastic noise is added to 
the input signal. 

Stochastic resonance has been shown to amplify signals
generated by linear models, such as sine waves. Recently, it has been
shown \cite{collinspre} that the same basic effect can be seen with
aperiodic (stochastic) input signals if the means of measurement is
appropriately modified. Since the spectrum of a random signal is not
discrete, the SNR must be replaced in this case with a power norm
sensitive to shape-matching and/or signal correlation.

Our present interest in stochastic resonance is somewhat
different. Our goal is in maximizing the amount of state information
carried by the spike train. We pose the question 
whether the quality of state information from a
{\it deterministic} signal (as opposed to the stochastic realization
used in \cite{collinspre}) can be improved by injecting random noise into the
process. In particular, we are interested in the case where the input
signal is chaotic. To determine the quality of state information
contained in the spike train, we measure the ability to predict the
spike train from its own history, using nonlinear prediction
techniques. We will show examples in which the interspike interval
prediction error of the output signal decreases (predictability
increases) with increasing noise added to the input signal. Showing
that nonlinear predictability is enhanced by adding noise is analogous
to the enhanced SNR shown in stochastic resonance studies.

The filters we will use to create spike trains capable of carrying
low-dimensional deterministic state information are based on the
well-known FitzHugh-Nagumo equation \cite{FHN}. The two dimensional
system FHN2 
\begin{eqnarray}
\eps \dot{v}& = & -v(v-0.5)(v-1) - w + S \nonumber\\
\dot{w} & = & v-w-b \label{eqfhn2}
\end{eqnarray}
is a simple differential equation that exhibits a fast spike followed
by a refractory period. There is an
equilibrium at $(v,w) = (v_0, v_0-b)$, where $v_0$ is a real-valued root of
$v_0(v_0-0.5)(v_0-1) = S+b-v_0$. A stability check of the
equilibrium $v_0$ shows the existence of a supercritical Hopf
bifurcation for  $S_H =v_H(v_H-0.5)(v_H-1)+v_H-b$, where $v_H =
0.5-\sqrt{3-12\eps}/6$. 
Therefore, if $b, \eps$ are fixed and the bifurcation parameter $S$ is
increased, the system undergoes a Hopf bifurcation at $S_H$,
 resulting in a periodic orbit of the system encircling the formerly stable
equilibrium. The periodic orbit is  manifested in rhythmical spiking
by the variable $v$. For example, setting $b=0.15, \eps = 0.005$, there
is a Hopf bifurcation point at $S_H\approx 0.112331\ldots$. For 
$S<S_H$, the system is quiescent; the equilibrium is
stable. For $S>S_H$,  the system spikes at a rate of
approximately 1 Hz.

Now consider FHN2 as a nonlinear filter by substituting for the
constant $S$ in (\ref{eqfhn2}) a
signal $S(t)$ from another system. Fig.~1 shows a plot of the variable
$v$ from (\ref{eqfhn2}) where $S$ has been replaced by a signal from
the R\"ossler system \cite{rossler}
\begin{eqnarray}
\dot{x} &=& \tau(-y-z)\nonumber\\
\dot{y} &=& \tau(x + ay) \nonumber\\
\dot{z} &=& \tau(b + (x - c)z) \label{eqross}
\end{eqnarray}
where the standard parameters are set to $ a = 0.36, b = 0.4, c = 4.5$, 
and $\tau = 0.5$ causes the trajectory to run at half speed.
The input signal, also plotted in Fig.~1, is 
$S(t) = 0.09+0.013x(t)$, where 
$x(t)$ is the $x$ variable of (\ref{eqross}). The bias of the signal
is $\langle S(t)\rangle = 0.093$, and its root mean square amplitude is
$\sqrt{\langle(S(t) - \langle S(t) \rangle)^2\rangle} = 0.035$. Fig.~2
demonstrates the threshold-crossing detection capability of
FHN2. When the 
peak height of $S(t)$ is greater than $\approx 0.15088$, FHN2 fires a
burst of spikes. Note that this threshold is significantly higher
than the Hopf bifurcation value $S_H$, which would be the threshold
in the limit of an $S(t)$ which oscillates infinitely slowly.

Although we expect the spike sequences to carry state information of
the R\"ossler system, because of its threshold detection behavior we
do not expect it to carry enough to reconstruct the entire
attractor. On the other hand, the benefit of a TC filter 
is that noise can in some cases enhance the reconstruction quality, as 
measured by prediction error. As in studies of stochastic
resonance, we will add white noise to the input signal of the filter
(in this case, the FHN2 spike generator).  The equation with 
noise term is
\begin{eqnarray}
\eps \dot{v}& = & -v(v-0.5)(v-1) - w + S(t) + \xi(t) \nonumber\\
\dot{w} & = & v-w-b, \label{eqfhn2n}
\end{eqnarray}
where $\eps=0.005$, $b=0.15$,  and $\xi(t)$ is Gaussian
white noise with zero mean and autocorrelation
$\langle\xi(t)\xi(s)\rangle = 2D\delta(t-s)$. For small values of the
noise level $D$ (including all those considered here), the variable
$v$ exhibits a clearly distinguishable spiking behavior, often in
bursts of more than one spike, as in Fig.~1. For analysis 
purposes, we found it more convenient to collect series 
of interburst intervals, each defined to be the elapsed time between
the final spike of one burst and the first spike of the next burst.
After using (\ref{eqfhn2n}) to make a series of 1024 interburst
intervals  \cite{MP}, we used a standard nonlinear prediction algorithm to
measure the level of determinism in the series.
The fact that state information from a deterministic system is
contained in a spike train, even when the spike train is chaotic, can
be detected by measuring the nonlinear predictability of the
interburst intervals. If it can be shown that the ISI series is
predictable ``beyond the power spectrum'', that is, if there is
predictability beyond that which is guaranteed by linear
autocorrelation, then there is evidence of nonlinear dynamics in the
series. 

The prediction algorithm works as follows.  Given an ISI vector
$V_0=(t_{i_0}, \ldots, t_{i_0-m+1})$, the 1\% of other reconstructed
vectors $V_k$ that are nearest to $V_0$ are collected, omitting
vectors $V_k$ close in time.
The ISI for some number $h$ of steps ahead are
averaged for all $k$ to make a prediction.  That is, the average
$p_{i_0}=\langle t_{i_k+h}\rangle_k$ is used to approximate the future
interval $t_{i_0+h}$.  The difference $p-t_{i_0+h}$ is the $h$-step
prediction
error at step $i_0$.  We could instead use the series mean $m$ to
predict at each step; this $h$-step prediction error is $m-t_{i_0+h}$.
The ratio of the root mean square errors of the two possibilities (the
nonlinear prediction algorithm and the constant prediction of the
mean) gives the normalized prediction error
NPE = $ \langle (p_{i_0}-t_{i_0+h})^2\rangle^{1/2}/
\langle (m-t_{i_0+h})^2\rangle^{1/2}$
where the averages are taken over the entire series.
The normalized prediction error is a measure of the (out-of-sample)
predictibility of the ISI series.  A value of NPE less
than $1$ means that there is linear or nonlinear predictability in the
series beyond the baseline prediction of the series mean.

The results of the predictability of the interburst interval series from
(\ref{eqfhn2n}) are shown in Fig.~3. For these parameter settings,
unlike those for Fig.~1, no spikes occurs in the absence of noise. As
the noise power $D$ is increased from zero, spikes begin to occur for
very small noise levels, although the interburst series show no
predictability (NPE $\approx 1$) until $D$ is raised beyond 
$10^{-11}$. The
prediction error then drops to a minimum and raises again when the
noise becomes large enough to swamp the system. The clearly noticeable
improvement in predictability due to extremely small noise input is 
essentially a stochastic resonance 
effect. These results show evidence of nonlinear
determinism, since Gaussian-scaled surrogate series \cite{surr}
created from all burst series considered in Fig.~3 have NPE $\approx 1$.

 A slight alteration in the
FitzHugh-Nagumo equations yields a nonlinear filter that acts as an
integrate-and-fire processor. Define the system FHN3 by
\begin{eqnarray}
\dot{u} &=& -au-cw+S(t) \nonumber\\
\eps\dot{v} &=& -v(v-0.5)(v-1)+u-dw \nonumber\\
\dot{w} &=& v^2-w-b. \label{eqfhn3}
\end{eqnarray}
This system is similar to FHN2 in that if $S(t)$ is set to be a
constant parameter $S$, there is a Hopf bifurcation as
$S$ is increased. Setting parameters $a = 0.1, b =
0.15, c = 0.5, d = 0.5, \eps = 0.005$, the bifurcation point is
$S_H\approx -0.059$. Its success as an information processor is shown in
Fig.~4. As with FHN2, we replace the parameter $S$ with an input
signal $S(t)$ from the R\"ossler attractor. The signal is
$S(t)=0.0023x(t)-0.04$, which corresponds to bias
$\langle S(t)\rangle = -0.04$ and  root mean square amplitude 
$\sqrt{\langle(S(t) - \langle S(t) \rangle)^2\rangle} =
0.006$. Comparing with FHN2 in Fig. 1, 
we see a marked difference in the way FHN3 processes the input signal.
Fig.~5(a) shows a three-dimensional plot of
the vectors $(t_i, t_{i-1}, t_{i-2})$, where $t_i = T_i-T_{i-1}$ is
the time interval between spikes of the $v$ variable of FHN3. 
Fig.~5(b) shows a similar reconstruction where the
input signal $S(t)$ is the $x$-coordinate from the Lorenz equations
\cite{lorenz} 
$\dot{x} = \tau(\alpha ( y - x )), 
\dot{y} = \tau(\rho x - y - xz), 
\dot{z} = \tau(-\beta z + xz)$,
where the parameters are set to the standard values 
$\alpha = 10, \rho = 28 , \beta = 8/3$, and $\tau = 0.01$.
Apparently, the interspike intervals recovered from (\ref{eqfhn3}) do
an effective job of reconstructing the chaotic attractor which
produced the input signal $S(t)$, for both the R\"ossler and Lorenz
examples. Nonlinear prediction on a length 1024 series of spikes
created as in Fig.~4 yields NPE $= 0.1$. This very low
NPE supports the visual indication in Fig.~5(a) of a faithful
 reconstruction of the underlying R\"ossler attractor.
This is similar to the mechanism that was studied in the generic
integrate-and-fire model of \cite{sauer}, where firing times $T_i$ were
generated recursively by 
\begin{equation}  \label{eqn1}
\int_{T_{i}}^{T_{i+1}} S(t)dt = \Theta
 \end{equation}
for a fixed threshold $\Theta$. Theoretical reconstruction results
for spike trains generated by model (\ref{eqn1}) are discussed in
 \cite{sauer}. 

Creating spikes using FHN2 or FHN3 means imposing a type of highly nonlinear
filter on the attractor signal, a filter which edits out amplitude
information (since the spike waveforms are essentially alike) and
converts the information entirely to event timings. Our purpose is to
gain insight into the data processing methods used in systems which
communicate through spike timings, as is conjectured for certain
neural systems \cite{neur}. We have shown by example that noise may be
useful for this communication, in that it can amplify transmission of
deterministic, nonlinear state information as measured by nonlinear
prediction error. For the latter spike generation model (FHN3),
we have the possibility of complete reconstruction of attractor
states.

\acknowledgments

The research of T.S. was supported in part by the National
Science Foundation (Computational Mathematics and Physics programs).

\begin{figure}
   \caption{The solid curve is the variable $v$ of FHN2, the 
FitzHugh-Nagumo equation (\ref{eqfhn2}) with $S$
replaced by $S(t) = 0.09+0.013x(t)$, where $x(t)$ is a solution of the
R\"ossler system (\ref{eqross}). The dashed curve is $S(t)$. When a
peak of $S(t)$ is greater than $\approx 0.15$, a burst is triggered in
FHN2.}
   \label{fig1}
\end{figure}

\begin{figure}
   \caption{Peak heights of the signal $S(t)$ from Fig.~1 graphed
   versus time. The height is plotted as an asterisk if it triggers a
   burst from FHN2; as an open circle if not. All of the asterisks lie
   above all of the open circles, signifying precise threshold
   detection by FHN2.} 
\label{fig2}
\end{figure}

\begin{figure}
   \caption{Normalized prediction error of spike trains generated by
   (\ref{eqfhn2n}), where $S(t)$ is a signal formed using the R\"ossler
   $x$-variable of (\ref{eqross}) with bias $0.075$ and rms amplitude
   $0.020$ (open circles) or $0.023$ (asterisks). As the input noise
   power $D$ increases, the NPE displays a minimum, 
   corresponding to maximum information transfer. Each plotted point
   is an average over 5 noise realizations; standard error is less
   than $0.02$ for each.}
   \label{fig3}
\end{figure}

\begin{figure}
   \caption{The solid curve is the variable $v$ of FHN3, equation
    (\ref{eqfhn3}), with $S(t)=0.0023x(t)-0.04$. The dashed curve is
    $x(t)$, the $x$-variable of the R\"ossler attractor (\ref{eqross}). A
    plot of $3$-tuples of interspike intervals from this equation is
    shown in Fig.~5a.}
    \label{fig4}
\end{figure}

\begin{figure}
    \caption{Interspike interval reconstructions of (a) the
   R\"ossler-FHN3 intervals from Fig.~4 (b) Lorenz-FHN3 intervals from
   (\ref{eqfhn3}) with $S(t) = .0005x(t)-0.04$, where $x(t)$ is the
   $x$-variable of the Lorenz system. In (b), fewer points are
   plotted, and they are connected with line segments.}
   \label{fig5}
\end{figure}


\begin{references}

\bibitem{PCFS} N. Packard, J. Crutchfield, J.~D. Farmer, and R. Shaw,
Phys. Rev. Lett., {\bf 45}, 712 (1980).
J.~C. Roux, A. Rossi, S. Bachelart, C. Vidal, Phys.
Lett. {\bf A77}, 391 (1980). J.~C. Roux, H. Swinney, in: {\it
Nonlinear Phenomena in Chemical Dynamics}, eds. C. Vidal, A. Pacault.
Springer-Verlag, Berlin (1981).

\bibitem{takens} F. Takens, Lecture Notes in Math. 898,
Springer-Verlag (1981). 


\bibitem{saueryc} T. Sauer, J.~A. Yorke, M. Casdagli, J. Stat. Phys.
{\bf 65} 579-616 (1991). 

\bibitem{sauer} T. Sauer, Phys. Rev. Lett. {\bf 72}, 3911
(1994). T. Sauer, in
Nonlinear Dynamics and Time Series, eds. C. Cutler, D. Kaplan.
Fields Institute Publications, Amer. Math. Soc. (1996).


\bibitem{CS} R. Castro, T. Sauer, Phys. Rev. E {\bf 55} (1997).

\bibitem{andre} D. Racicot, A. Longtin, Physica D (in press).

\bibitem{FHN} R. FitzHugh, In: Biological Engineering,
Ed. by H.P. Schwann. McGraw-Hill, New York (1962). J. Nagumo,
S. Arimoto, S. Yoshizawa, Proc. IRE {\bf 50}, 2061 (1962).

\bibitem{sr} K. Weisenfeld, F. Moss, Nature {\bf 373}, 33 (1995).
F. Moss, A. Bulsara, M.F. Schlesinger, Eds. {\it Proc. of
NATO Advanced Research Workshop on Stochastic Resonance in Physics and
Biology}, J. Stat. Phys. {\bf 70} (1993). A.
Longtin, A. Bulsara, F. Moss, Phys. Rev. Lett. {\bf 67}, 656 (1991).

\bibitem{collinspre} J.J. Collins, C.C. Chow, T.T. Imhoff,
Phys. Rev. E {\bf 52}, R3321 (1995).

\bibitem{rossler} O.~E. R\"ossler, Physics Letters {\bf 57A}, 397
(1976).

\bibitem{MP} The method of R. Manella and V. Palleschi, Phys. Rev. A
{\bf 40}, 3381 (1989) 
was used for integrating (\ref{eqfhn2n}), with a step size of $0.001$.

\bibitem{surr} J. Theiler, S. Eubank, A. Longtin, B. Galdrakian,
J.D. Farmer, Physica D {\bf 58}, 77 (1992).

\bibitem{lorenz} E. Lorenz, J. Atmos. Sci. {\bf 20} 131 (1963).

\bibitem{neur} W. Bialek, F. Rieke, R. R. de Ruyter van Steveninck,
D. Warland, Science {\bf 252}, 1854 (1991). L. Abbott,
Quart. Rev. Biophys. {\bf 27}, 291 (1994). E. Vaadia, I. Haalman,
M. Abeles, H. Bergman, Y. Prut, H. Slovin, A.M.H.J. Aertsen, Nature
{\bf 373}, 5151 (1995). Z. Mainen, T. J. Sejnowski, Science
 {\bf 268}, 1503 (1995).

\end{references}
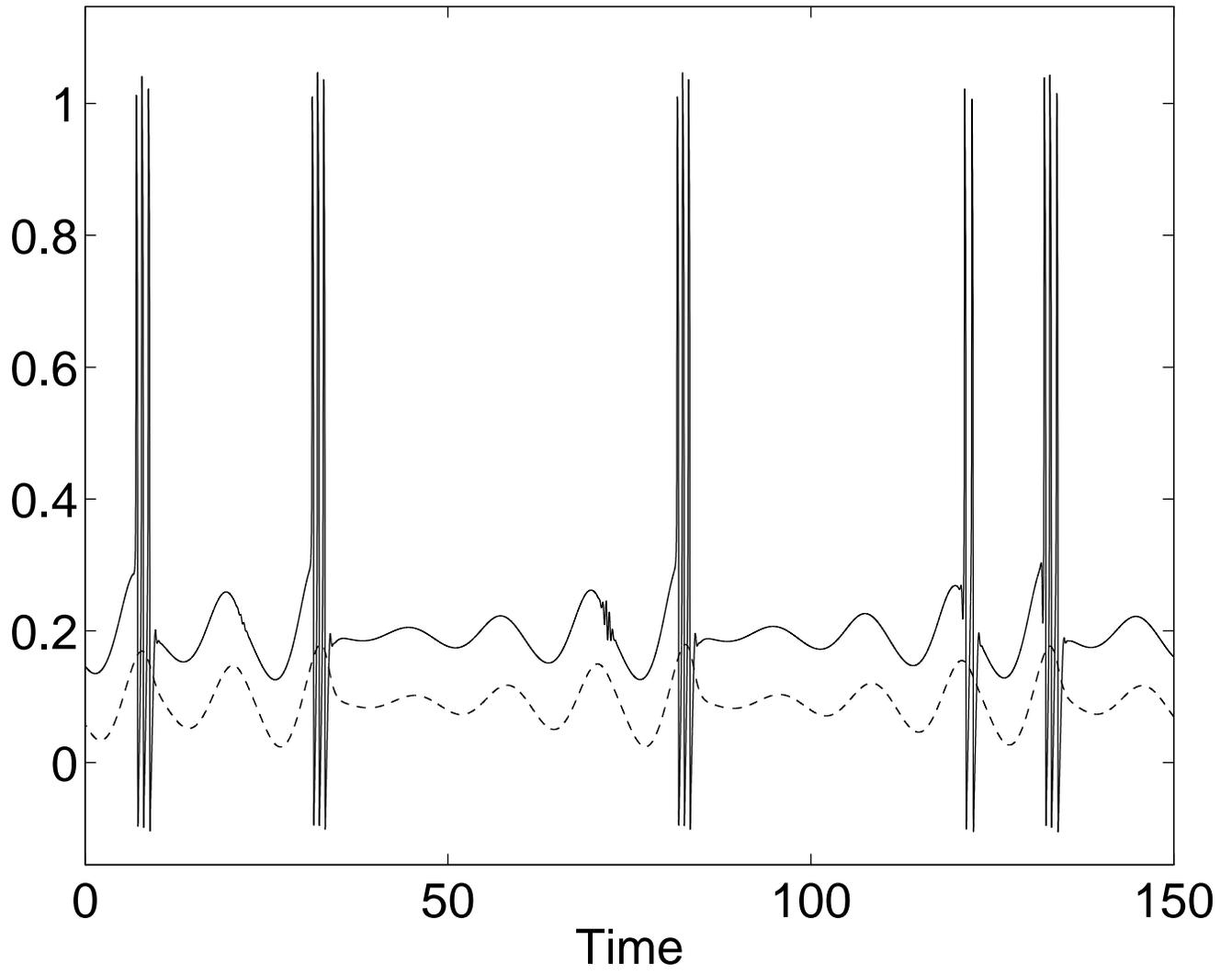
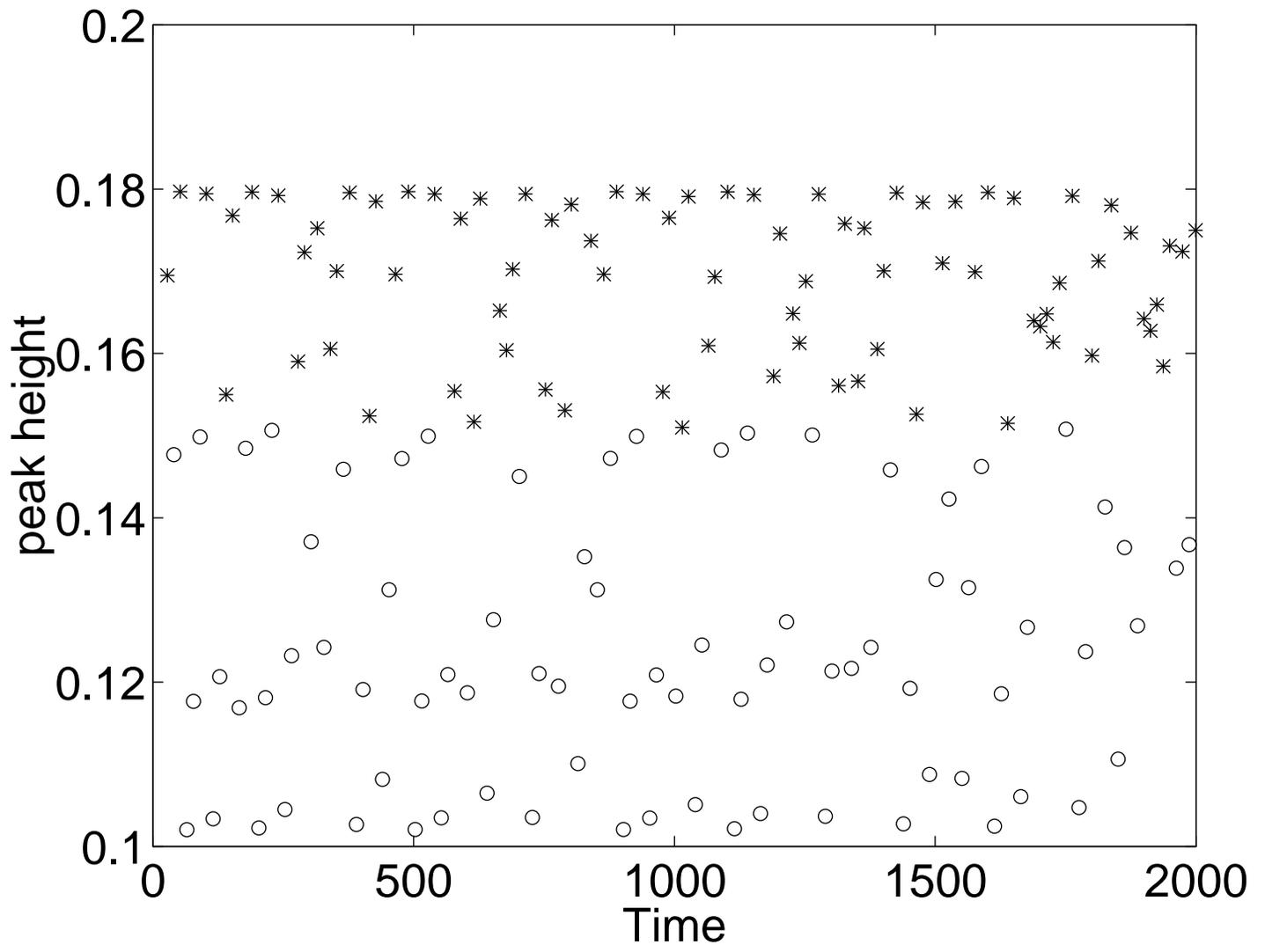
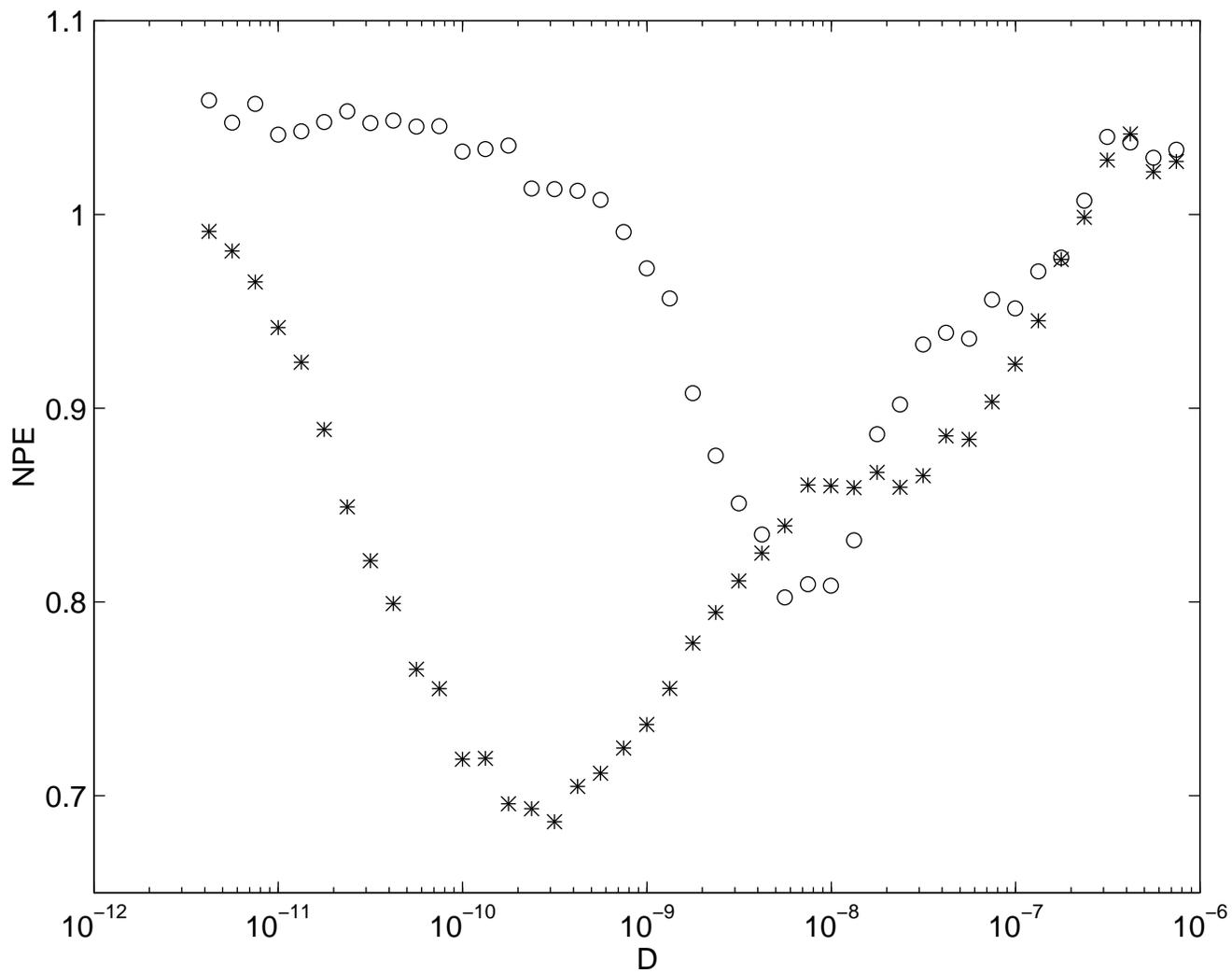
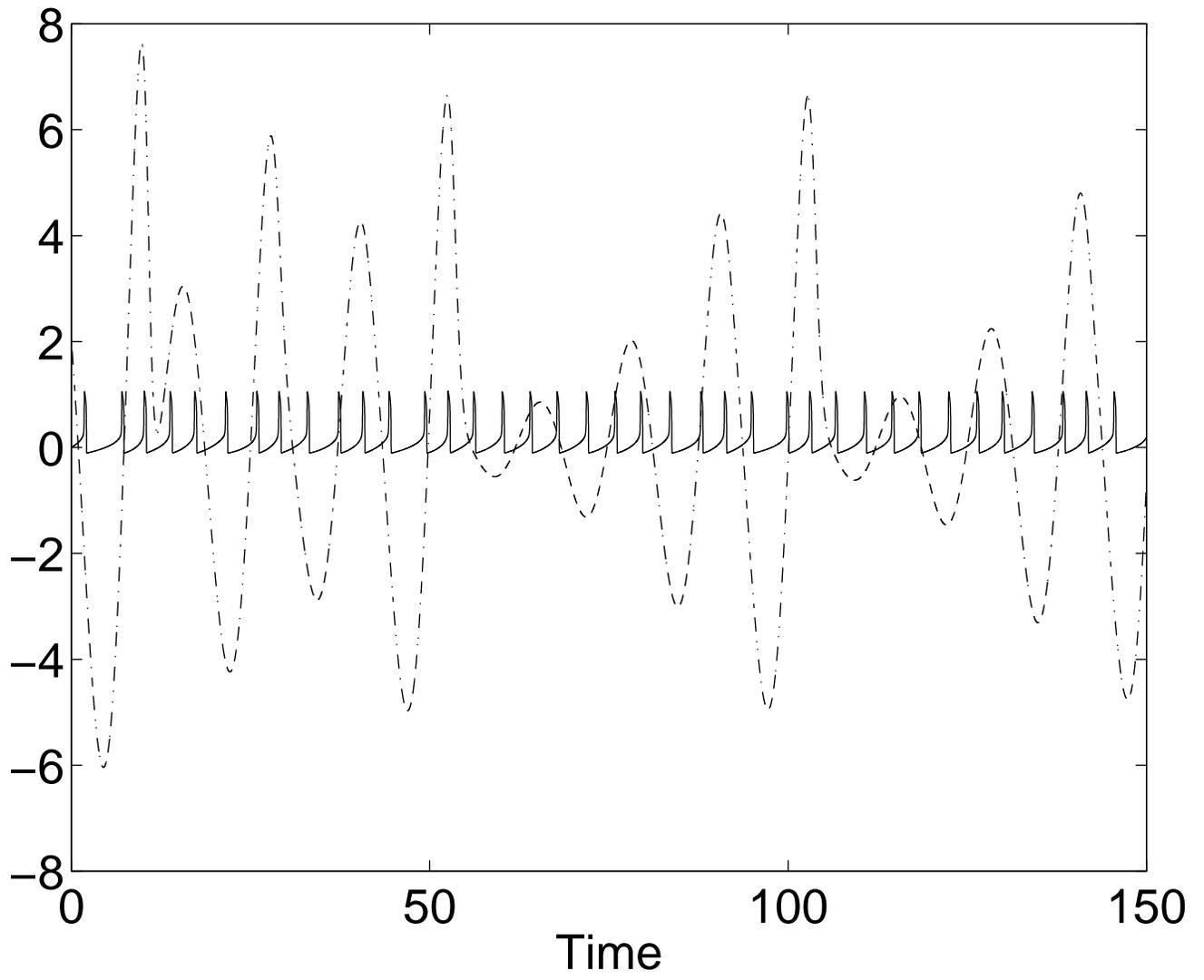
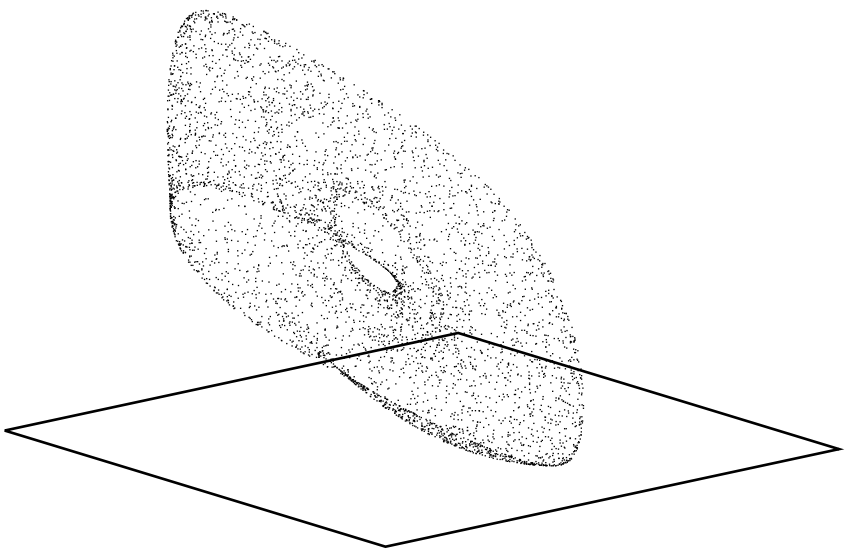
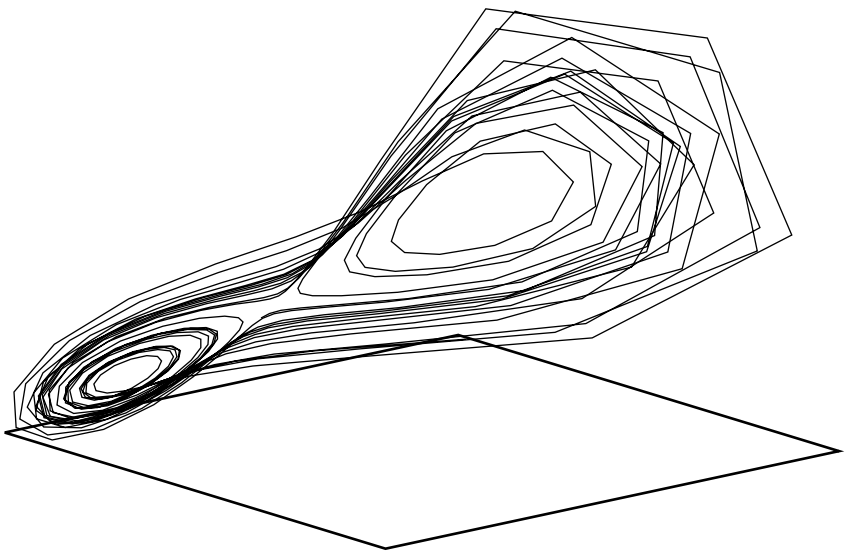
\end{document}